# STUDYING THE FEASIBILITY AND IMPORTANCE OF SOFTWARE TESTING: AN ANALYSIS

Dr.S.S.Riaz Ahamed

Principal, Sathak Institute of Technology, Ramanathapuram,India.

Email:ssriaz@ieee.org, ssriaz@yahoo.com

**ABSTRACT**

Software testing is a critical element of software quality assurance and represents the ultimate review of specification, design and coding. Software testing is the process of testing the functionality and correctness of software by running it. Software testing is usually performed for one of two reasons: defect detection, and reliability estimation. The problem of applying software testing to defect detection is that software can only suggest the presence of flaws, not their absence (unless the testing is exhaustive). The problem of applying software testing to reliability estimation is that the input distribution used for selecting test cases may be flawed. The key to software testing is trying to find the modes of failure - something that requires exhaustively testing the code on all possible inputs. Software Testing, depending on the testing method employed, can be implemented at any time in the development process.

## 1 INTRODUCTION

Testing is a set of activities that could be planned ahead and conducted systematically. The main objective of testing is to find an error by executing a program. The objective of testing is to check whether the designed software meets the customer specification. The Testing should fulfill the following criteria:

- Test should begin at the module level and work "outward" toward the integration of the entire computer based system.
- Test should be appropriate and different for different points in time.
- An independent test group should exist and should conduct testing for different software development projects.
- Though testing and debugging are different activities, debugging must be accommodated in any testing activity.
- A methodology for software testing must accommodate low-level tests that are necessary to verify that a small source code segment has been correctly implemented as well as high-level tests that validate major system functions against customer requirements. It should indeed provide guidance for the practitioner and a set of milestones for the manger.

## 2 TEST PLAN

For performing any activity, planning is done and similarly testing commences with a test plan. A test plan is a general document for the entire project that defines the scope, approach to be taken, and the schedule of testing as well as identifies the test items for the entire testing process and the personnel responsible for the different activities of testing. The inputs for the test plan are viz, project plan, requirements documents and system design document. An ideal test plan should contain the following:
- Test unit specification
- Features to be tested
- Approach to testing
- Test deliverables
- Schedule
- Personnel allocation

**Testability**

Software testability is the process of testing how adequately a particular set of tests will cover the product. This enables the individuals charged with testing to design effective test cases more easily. As per the definition of James Bach, Software testability is simply how a computer program can be tested. It is clear that testing is not small process and it is profoundly difficult, it pays to know what can be done to streamline it.

The attributes of good test are
- ➢ It should have a high probability of finding an error: The tester must understand the software and attempt to develop a mental picture of the failure possibilities.
- ➢ It is not redundant: Since the testing time and resources are limited, there is no point of conducting the test of same procedure as like the other. Every test should have different purpose.
- ➢ It should be best of breed.
- ➢ It should be neither simple nor complex. Although it is sometimes possible to combine a series of tests into one test case, the possible side effects associated with this approach may mask errors.

## 3 DEVELOPMENT TESTING

Development testing includes verification and validation (V & V). Verification refers to the set of activities that ensure that software correctly implements a specific function. Validation refers to a different set of activities that ensure that the software has been built is traceable to customer requirements.
Verification: "Are we building the product right"?
Validation: "Are we building the right product"? Some of the activities of verification and validation are part of the software quality assurance that include formal technical reviews, quality and configuration audits, performance monitoring, simulation, feasibility study, document review, database review, algorithm analysis, development testing, qualification testing, and installation testing.

Tom Gilb suggests that the following issues must be answered if testing strategy is to be implemented:
1. Specify product requirement in quantifiable manner long before testing commences. Although the overriding objective of testing is to find errors, a good testing strategy also assesses other quality characteristics such as portability, maintainability, and usability. These must be specified in a way that is measurable so that testing results are unambiguous.
2. State testing objectives explicitly. The specific objectives of testing should be stated in measurable terms. For example, test effectiveness, test coverage, meantime to failure, the cost to find and fix defects, remaining defect density or frequency of occurrence, and test work hours per regression test should all be stated within the test plan.
3. Understand the users of the software and develop a profile for each user category. The description of the interaction scenario for each class of user can reduce overall testing effort by focusing testing on actual use of the product.
4. Develop a testing plan that emphasizes "rapid cycle testing": Tom Gilb recommends that a software engineering team "learn to test in rapid cycles (2% of project effort) of customer-useful, at least field 'trialable', increments of functionality and/or quality improvement." The feedback generated from these rapid cycle tests can be u sed to control quality levels and the corresponding test strategies.
5. Build "robust" software that is designed to test itself. Software should be designed in a manner that uses antibugging techniques. That is, software should be capable of diagnosing certain classes of errors. In addition, the design should accommodate automated testing and regression testing.
6. Use effective formal control reviews as a filter prior to testing. Formal technical reviews can uncover inconsistencies, omissions, and outright errors in the testing approach. This saves time and also improves product quality.
7. Develop a continuous improvement approach for testing process. The test strategy should be measured. The metrics collected during testing should be used as part of a statistical process control approach for software testing.

## 4 GENERAL TESTING METHODS

Software bugs are a fact of life. No matter how hard we try, the reality is that even the best programmers can't write error-free code all the time. On average, even well-written programs have one to three bugs for every 100 statements. It is estimated that testing to find those bugs consumes half the labor involved in producing a working program. Statistics like these explain why so much attention is focused on making testing more effective.

Traditionally, there are two main approaches to testing software: black-box (or functional) testing and white-box (or structural) testing. Black-box and white-box testing should be performed together for every application so that the software will live up to users' expectations on every level.

**White-Box testing**
White-box testing strategies include designing tests such that every line of source code is executed at least once, or requiring every function to be individually tested. Very few white-box tests can be done without modifying the program, changing values to force different execution paths, or to generate a full range of inputs to test a particular function. Traditionally, this modification has been done using interactive debuggers, or by actually changing the source code. While this may be adequate for small programs, it does not scale well to larger applications. Traditional debuggers greatly affect the timing, sometimes enough so that a large application will not run without major modifications. Changing the source code is also unwieldy on a large program that runs in a test bed environment.

The intention in white box testing is to ensure that all possible feasible flow of control paths through a subprogram are traversed whilst the software is under test. This is not the same as saying that all statements in the subprogram will be executed as it is possible for all statements to be executed but for not all of the possible paths to be traversed. However, the converse is true; if all possible paths through a subprogram are traversed then all statements in the subprogram will necessarily be executed.

White-box testing requires visibility into the executable to determine what to test. It also requires a method to determine the outcome of the test. The ability to output values from within the application in the most noninvasive manner possible is a necessary capability in any white-box testing tool. **W**hite-box testing is quickly gaining acceptance as a method of testing code. One way to examine the merits of white-box testing is to look at it in terms of its more commonly practiced counterpart: black-box testing. White-box testing is a method of verifying that code has been constructed properly and has no hidden weaknesses.

The goal of white-box testing is to execute every branch of code under different input conditions to make sure the code does not exhibit abnormal behavior. You are flooding the application with inputs to expose any construction problems that will cause the application to crash. You should try to call every public method in every class with different arguments to verify that the methods do not exhibit uncaught runtime exceptions under these circumstances.

White-box testing is already an important part of the development cycle in some companies. Developers who are not yet performing white-box testing have probably been scared off by one or more of the white-box testing myths floating around the development community. In this article, I will dispel some of the most common myths about white-box testing.

*Myth:* White-box testing is redundant. I can get the same results from black-box testing.

*Fact:* Black-box testing, while important to software quality, can't test your code's structure the way white-box testing can. There is a fundamental difference in the purposes of white-box and black-box testing. Black-box testing is functionality testing. Developers perform black-box testing to see whether the application behaves according to specifications. White-box testing is construction testing; it verifies that each class contains no hidden flaws.

**Timing**
A fundamental measurement in any system is time. Many system requirements are specified in terms of time, particularly response time, and a relatively large amount of testing is focused on timing. The necessary primitive for white-box testing of timing is a precise clock and the ability to take times at various points in the program while affecting the program's execution as little as possible.

**Fault Injection**
In any operational environment, faults occur and must be handled correctly. Even the simplest programs must be able to recover from mistakes: mistyped user inputs, disk-full conditions, and such.

This raises a problem: How do you induce faults to test how fault-tolerant a system is? For example, if the system is supposed to be able to recover from disk errors, how do you test for that?
In the past, users had to modify the source code to simulate the various faults, explicitly changing the application to make it take error paths. Given that a system may need to tolerate hundreds of faults, this can be onerous. Also, given the amount of work involved, it is unlikely to be repeated when modifications are made to the software.

With a white-box testing tool, fault injection becomes simple and can be repeated as part of normal regression testing.

**Test Coverage**
Test coverage is an important component of white-box testing. The goal is to try to execute (that is, test) all lines in an application at least once.

Because white-box testing tools can individually or collectively instrument source lines, it is straightforward to determine which lines in a host program have or have not been executed without modifying source. Aprobe can reference and change variables in the application, so it can easily support the test coverage effort.

However, what about exception handlers/catchers? How do you get the code in an exception handler/catcher to be executed? Using probes like this, you can easily force different execution paths to ensure execution of all lines.

**Assertions**
A common white-box technique for increasing software quality is to add assertions to the code. These assertions typically abort the program as soon as a fatal error is detected, so the error does not propagate and become more difficult to diagnose. Usually, assertions are added to source code and turned on/off at compilation time. However, adding assertions to the executable is easier. Any number of assertions can be coded and added to the executable on demand. They can be added or removed on each run, without having to recompile the code.

White-box testing is a powerful software testing technique. However, it can be cumbersome when source code modifications are needed. Source-level patching of the executable gives you valuable insight into the way a program works. Having the ability to force errors and stress the application allows for more thorough testing. Because testing can proceed without waiting for source code changes, more extensive testing can occur in a shorter time frame. When used in combination with black-box testing, white-box tools allow more thorough and comprehensive testing of an application. The result is an increase in the quality of the software

White-box test design allows one to peek inside the "box", and it focuses specifically on using internal knowledge of the software to guide the selection of test data. Synonyms for white-box include: structural, glass-box and clear-box.

**Black-Box Testing**
Black-box testing tests whether an application actually functions as it is intended to function. This type of testing is performed by comparing an application's actual functionality with the intended functionality described in the application's specification document. In terms of building inspection, black-box testing might involve checking that every home element included in the blueprint is actually present and meets the specifications set forth in the blueprint. There are two types of black-box testing that can be applied to Web applications:

> ➢ Testing critical paths' functionality (i.e., testing certain Site functionality by testing if associated paths through the site contain errors).
> ➢ Testing whether or not all appropriate pages contain certain invariable elements.

Black-box test design treats the system as a "black-box", so it doesn't explicitly use knowledge of the internal structure. Black-box test design is usually described as focusing on testing functional requirements. Synonyms for black-box include: behavioral, functional, opaque-box, and closed-box.

In black-box testing, developers first set up a test plan that will test all the ways users will interact with the application. These interactions can range from simply opening the program to selecting lower-level menu options. As soon as the test plan is in place, the developer should come up with inputs that will actually test the program.

In black-box testing, software is exercised over a full range of inputs and the outputs are observed for correctness. How those outputs are achieved -- or what is inside the box -- doesn't matter.

Although black-box testing has many advantages, by itself it is not sufficient.

> ➢ First, real-life systems have too many different kinds of inputs, resulting in a combinatorial explosion of test cases.

- ➢ Second, the correct operation of the program may not be a measurable output.
- ➢ Third, it is impossible to determine whether portions of the code have even been executed by black-box testing. Code that has not been executed during testing is a sleeping bomb in any software package. Certainly, code that has not been executed has not been tested.
- ➢ Finally, and perhaps most convincingly, empirical evidence shows that black-box testing alone does not uncover as many errors as a combination of testing methods does

**Black-Box and White-Box testing Methods**
While black-box and white-box are terms that are still in popular use, many people prefer the terms "behavioral" and "structural". Behavioral test design is slightly different from black-box test design because the use of internal knowledge isn't strictly forbidden, but it's still discouraged. In practice, it hasn't proven useful to use a single test design method. One has to use a mixture of different methods so that they aren't hindered by the limitations of a particular one. Some call this "gray-box" or "translucent-box" test design, but others wish we'd stop talking about boxes altogether.

It is important to understand that these methods are used during the test design phase, and their influence is hard to see in the tests once they're implemented. Note that any level of testing (unit testing, system testing, etc.) can use any test design methods. Unit testing is usually associated with structural test design, but this is because testers usually don't have well-defined requirements at the unit level to validate.

White box testing is concerned only with testing the software product, it cannot guarantee that the complete specification has been implemented. Black box testing is concerned only with testing the specification, it cannot guarantee that all parts of the implementation have been tested. Thus black box testing is testing against the specification and will discover *faults of omission*, indicating that part of the specification has not been fulfilled. White box testing is testing against the implementation and will discover *faults of commission*, indicating that part of the implementation is faulty. In order to fully test a software product both black and white box testing are required.
White box testing is much more expensive than black box testing. It requires the source code to be produced before the tests can be planned and is much more laborious in the determination of suitable input data and the determination if the software is or is not correct. The advice given is to start test planning with a black box test approach as soon as the specification is available. White box planning should commence as soon as all black box tests have been successfully passed, with the production of flowgraphs and determination of paths. The paths should then be checked against the black box test plan and any additional required test runs determined and applied.

The consequences of test failure at this stage may be very expensive. A failure of a white box test may result in a change which requires all black box testing to be repeated and the re-determination of the white box paths. The cheaper option is to regard the process of testing as one of *quality assurance* rather than *quality control*. The intention is that sufficient quality will be put into all previous design and production stages so that it can be expected that testing will confirm that there are very few faults present, *quality assurance*, rather than testing being relied upon to discover any faults in the software, *quality control*.

**A Software Testing Strategy**
Testing are done for various purposes based on the need, necessity and importance.
A complete testing strategy consists of testing at various points in the production process and can be described by the *test vee*

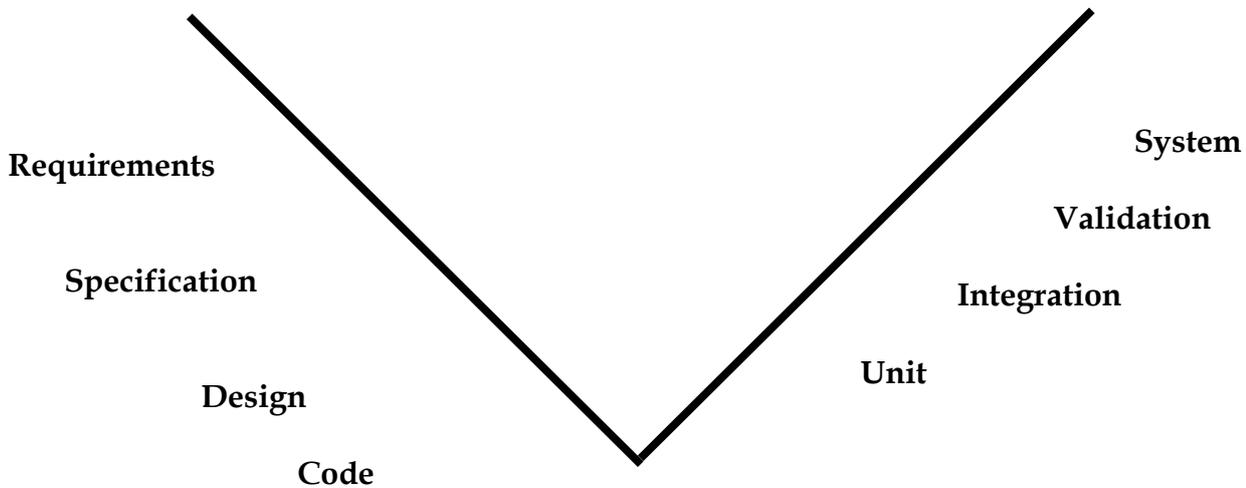

The left hand side of the V indicates the processes involved in the construction of the software, starting with the determination of requirements which are subsequently refined into a precise specification. The design phase in this model is taken to indicate the design of modules and the code phase the detailed design and construction of subprograms.

The right hand side of the V indicates the testing actions which correspond to each stage on the left side. *Unit testing* is concerned with the testing of the individual subprograms, *integration testing* with the assembly of the modules to produce the application, *validation testing* with ensuring that the application meets its specification and *system testing* that it serves its requirements and fits into its environment. Although this model presents these four actions as distinct stages, in practice the stages overlap to a considerable extent

**5 UNIT TESTING**
It is a testing process, which focuses verification effort on the smallest unit of software design- the module. Using the procedural design description as guide, important control paths are tested t uncover errors within the boundary of the module. The relative complexity of tests and uncovered errors is limited by the constrained scope established for unit testing. The unit test is normally whit e box oriented, and the step can be conducted in parallel for multiple modules.

The unit test makes the programmer satisfied that the software does what the programmer thinks it does.

I think many people confuse traditions with the actual definition of the tests. One might say, for example, "A unit test is testing an individual class in isolation," or "An acceptance test tests the entire program." That would be wrong. They are not the definitions, just traditions resulting from the forces that act upon you when you are doing testing. There is absolutely nothing wrong with the programmer writing a so-called unit test that tests the whole program, or with the customer defining a functional test that stubs out part of the system. It's up to the people doing it to weigh the costs and benefits, not to be arbitrarily constrained to tradition.

Tests of data flow across a module interface are required before any other test is initiated. If data do not enter and exit properly, all other tests are moot. Interface will have to be tested from the following angle:
1. Number of input parameters are equal to number of arguments?
2. Parameters and argument attributes match?
3. Parameter and argument units system match?
4. Number of arguments transmitted to called modules are equal to number of parameters?
5. Attributes of arguments transmitted to called modules equal to attributes of parameters?
6. Unit system of arguments transmitted to called modules equal to unit system of parameters?
7. Number attributes and order of arguments to built-in-functions correct?
8. Any references to parameters not associated with current point of entry?
9. Input only arguments altered?

10. Global variable definitions consistent across modules?
11. Constraints passed as arguments?

When a module performs external I/O, additional interface tests must be conducted. Again, from Myers
1. File attributes correct?
2. OPEN/CLOSE statements correct?
3. Format specification matches I/o statement?
4. Buffer size matches record size?
5. Files opened before use?
6. End-of-file conditions handled?
7. I/O errors handled?
8. Any textual errors in output information?

The local data structure for a module is a common source of errors. Test cases should be designed to uncover errors in the following categories:
1. improper or inconsistent typing
2. erroneous initialization or default values
3. incorrect(misspelled or truncated) variable names
4. inconsistent data types
5. underflow, overflow, and addressing exceptions

The impact of global data on a module should be ascertained (if possible) during unit testing. Selective testing of execution paths is an essential task during the unit test. Test cases should be designed to uncover errors due to erroneous computations, incorrect comparisons, or improper control flow. Basis path and loop testing are

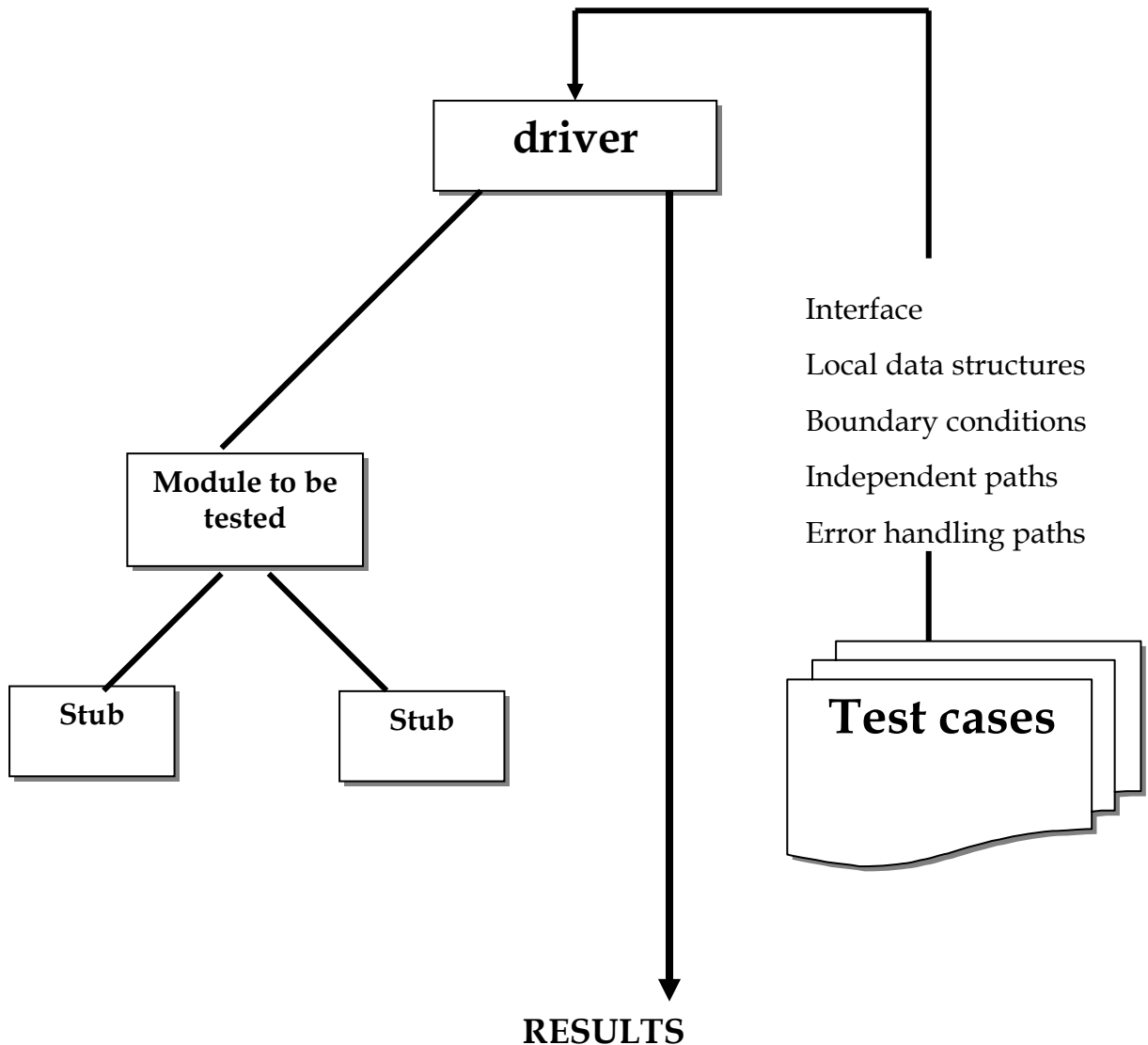

**RESULTS**

effective techniques for uncovering a broad array of path errors. Among the more common errors in computation are (1) misunderstood or incorrect arithmetic precedence; (2) mixed mode operations; (3) incorrect initialization; (4) precision inaccuracy; and (5) incorrect symbolic representation of an expression. Comparison and control flow are closely coupled to one another (i.e. change of flow frequently occurs after a comparison). Test cases should uncover errors such as (1) comparison of different data types; (2) incorrect logical operators or precedence; (3) expectation of equality when precision errors makes equality unlikely; (4) incorrect comparison or variables; (5) improper or inconsistent loop termination: (6) failure to exit when divergent iteration is encountered; and (7) improperly modified loop variables.

In most cases, it is construed that during design the errors are anticipated and visualized. When error handling is evaluated the following potential errors are tested:
1. Error description is unintelligible.
2. Error noted does not correspond to error encountered
3. Error conditions causes system intervention prior to error handling
4. Exception-condition processing is incorrect.
5. Error description does not provide enough information to assist in the location of the cause of the error.

Unit test is normally considered after the coding is done. Once the program is reading after having reviewed, syntax errors corrected, the case of unit test begins. Because a module is not a standalone program, driver and/or a stub software must be developed for each unit test. Unit testing is simplified when a module with high cohesion is designed. When only one function is addressed by a module, the number or test cases is reduced and errors can be more predicted and uncovered.

**Function Test**
A formal test conducted to determine whether or not a system satisfies its *acceptance criteria* and to enable the customer to determine whether or not to accept the System. A better name for Functional Test. is acceptance test.

Approach to testing can be undertaken in different ways. Test cases are decidedly mainly as per the specifications of the program and the internals of the program or module are insignificant. The functional testing is often called "black box testing" The basis for deciding test cases in functional testing is the requirements or specifications of the system or module. For the entire system, the test cases are designed from the requirements specification document for the system. For modules created during design, test cases for functional testing are decided from the module specifications produced during the design.

Though there are no formal rules for designing test cases for functional testing, there are a number of testing or heuristics that can be used to select test cases that have been by history found to be very successful in detecting errors. They are:

- Equivalence class partitioning
- Boundary value analysis
- Cause-Effect graphing

| UNIT TEST | FUNCTION TEST |
|---|---|
| It test the behavior of a single class | The acceptance test makes the customer satisfied that the software provides the business value that makes them willing to pay for it |
| It help development go faster as well as ensuring quality | |
| It belong to the developers | It test the entire system from end to end. Like Unit Test help development go faster as well as ensuring quality |
| | It is the responsibility of the customer |

**Integration Testing**
Integration testing can proceed in a number of different ways, which can be broadly characterised as *top down* or *bottom up*. In *top down integration testing* the high level control routines are tested first, possibly with the middle level control structures present only as *stubs*.

Top down testing can proceed in a *depth-first* or a *breadth-first* manner. For depth-first integration each module is tested in increasing detail, replacing more and more levels of detail with actual code rather than stubs. Alternatively breadth-first would proceed by refining all the modules at the same level of control throughout the application. In practice a combination of the two techniques would be used. At the initial stages all the modules might be only partly functional, possibly being implemented only to deal with non-erroneous data. These would be tested in breadth-first manner, but over a period of time each would be replaced with successive refinements which were closer to the full functionality. This allows depth-first testing of a module to be performed simultaneously with breadth-first testing of all the modules.

Steps:
1. Main control module used as the test driver, with stubs for all subordinate modules.
2. Replace stubs either depth first or breadth first.
3. Replace stubs one at a time.
4. Test after each module integrated.
5. Use regression testing (conducting all or some of the previous tests) to ensure new errors are not introduced.
6. Verifies major control and decision points early in design process.
7. Top level structure tested the most.

The other major category of integration testing is *bottom up integration testing* where an individual module is tested from a test harness. Once a set of individual modules have been tested they are then combined into a collection of modules, known as *builds*, which are then tested by a second test harness. This process can continue until the build consists of the entire application.

Steps:
1. Low level modules combined in clusters (builds) that perform specific software subfunctions.
2. Driver program developed to test.
3. Cluster is tested.
4. Driver programs removed and clusters combined, moving upwards in program structure.

**Validation Testing**
*Validation testing* is a concern, which overlaps with integration testing. Ensuring that the application fulfils its specification is a major criterion for the construction of an integration test. Validation testing also overlaps to a large extent with *system testing*, where the application is tested with respect to its typical working environment. Consequently for many processes no clear division between validation and system testing can be made. Specific tests which can be performed in either or both stages include the following.

> - *Regression testing*. Where this version of the software is tested with the automated test harnesses used with previous versions to ensure that the required features of the previous version are still working in the new version.
> - *Recovery testing*. Where the software is deliberately interrupted in a number of ways, for example taking its hard disc off line or even turning the computer off, to ensure that the appropriate techniques for restoring any lost data will function.
> - *Security testing*. Where unauthorised attempts to operate the software, or parts of it, are attempted. It might also include attempts to obtain access the data, or harm the software installation or even the system software. As with all types of security it is recognised that someone sufficiently determined will be able to obtain unauthorised access and the best that can be achieved is to make this process as difficult as possible.
> - *Stress testing*. Where abnormal demands are made upon the software by increasing the rate at which it is asked to accept data, or the rate at which it is asked to produce information. More complex tests may attempt to create very large data sets or cause the software to make excessive demands on the operating system.
> - *Performance testing*. Where the performance requirements, if any, are checked. These may include the size of the software when installed, the amount of main memory and/ or secondary storage it requires and the demands made of the operating system when running within normal limits or the response time.
> - *Usability testing*. The process of usability measurement was introduced in the previous chapter. Even if usability prototypes have been tested whilst the application was constructed, a validation test of the finished product will always be required.

> *Alpha and beta testing*. This is where the software is released to the actual end users. An initial release, the alpha release, might be made to selected users who would be expected to report bugs and other detailed observations back to the production team. Once the application has passed through the alpha phase a beta release, possibly incorporating changes necessitated by the alpha phase, can be made to a larger more representative set users, before the final release is made to all users.

The final process should be a *software audit* where the complete software project is checked to ensure that it meets production management requirements. This ensures that all required documentation has been produced, is in the correct format and is of acceptable quality. The purpose of this review is: firstly to assure the quality of the production process and by implication the product; and secondly to ensure that all is in order before the initial project construction phase concludes and the maintenance phase commences. A formal hand over from the development team at the end of the audit will mark the transition between the two phases.

**System testing**
The development of complex High Availability, High Throughput software systems is only part of an overall project - testing them thoroughly, both in terms of functionality and their ability to handle peak - or extreme - transaction rates, is a major task. Another problem is the availability of suitable hardware and software at the appropriate time in a project - and how easy it is to configure these items to generate different load profiles

**6 CONCLUSION**

Software testing is an important phase in the software development life cycle. It represents the ultimate review of specification, design and coding. The main objective for test case design is to derive a set of tests that can find out errors in the software. A thoroughly tested software maintains quality. A quality product is what every one wants and appreciates. Software Testing also provides an objective, independent view of the software to allow the business to appreciate and understand the risks at implementation of the software. Test techniques include, but are not limited to, the process of executing a program or application with the intent of finding software bugs. The scope of software testing often includes examination of code as well as execution of that code in various environments and conditions as well as examining the aspects of code: does it do what it is supposed to do and do what it needs to do. In the current culture of software development, a testing organization may be separate from the development team. There are various roles for testing team members.